\documentclass[dvips]{article}
\usepackage{spie}

\usepackage{times}
\usepackage{graphicx}
\usepackage{calc}

\title{Multi-wavelength visibility measurements of the red giant R~Doradus}

\author{A.~P.~Jacob\supit{a},
        T.~R.~Bedding\supit{a},
	J.~G.~Robertson\supit{a},
	J.~R.~Barton\supit{b},
	C.~A.~Haniff\supit{c},
	R.~G.~Marson\supit{d} \\and
        M.~Scholz\supit{e}  
\skiplinehalf
\supit{a}School of Physics, University of Sydney 2006, Australia\\
\supit{b}Anglo-Australian Observatory, PO Box 296, Epping, NSW, 2121,
        Australia \\ 
\supit{c}Astrophysics Group, Cavendish Laboratory, Madingley Road,
	Cambridge CB3 OHE, UK\\
\supit{d}NRAO, PO Box 0, Socorro, NM 87801, USA\\
\supit{e}Institut f. Theoretische Astrophysik der Universit{\"a}t
        Heidelberg,\\ Tiergartenstr. 15, 69121 Heidelberg, Germany
}

\authorinfo{Send correspondence to Tim Bedding (\tt
bedding@physics.usyd.edu)}

\begin{document}
  \maketitle

\begin{abstract}
We present visibility measurements of the nearby Mira-like star R~Doradus
taken over a wide range of wavelengths (650--990\,nm).  The observations
were made using MAPPIT (Masked APerture-Plane Interference Telescope), an
interferometer operating at the 3.9-m Anglo-Australian Telescope.  We used
a slit to mask the telescope aperture and prism to disperse the
interference pattern in wavelength.  We observed in R~Dor strong decreases
in visibility within the TiO absorption bands.  The results are in general
agreement with theory but differ in detail, suggesting that further work is
needed to refine the theoretical models.
\end{abstract}


\section{The MAPPIT Interferometer}
\label{sec.mappit}

MAPPIT (Masked APerture-Plane Interference Telescope) is an interferometer
operating at the 3.9-m Anglo-Australian Telescope\cite{Bed92,BRM94b,Mar97}.
Its main features are a one-dimensional aperture mask (either a slit or an
array of holes) and a prism to disperse the fringes in wavelength.  This
allows us to record interference fringes simultaneously over a wide
wavelength range.  Here we report multi-wavelength visibility measurement
of the Mira-like red giant star R~Doradus, which we compare with our recent
theoretical work on the wavelength dependence of angular diameters of
M~giants\cite{Jac2000,JBR2000}.

\begin{figure}
\centering
\includegraphics[bb=41 156 513 297,width=13.5cm]{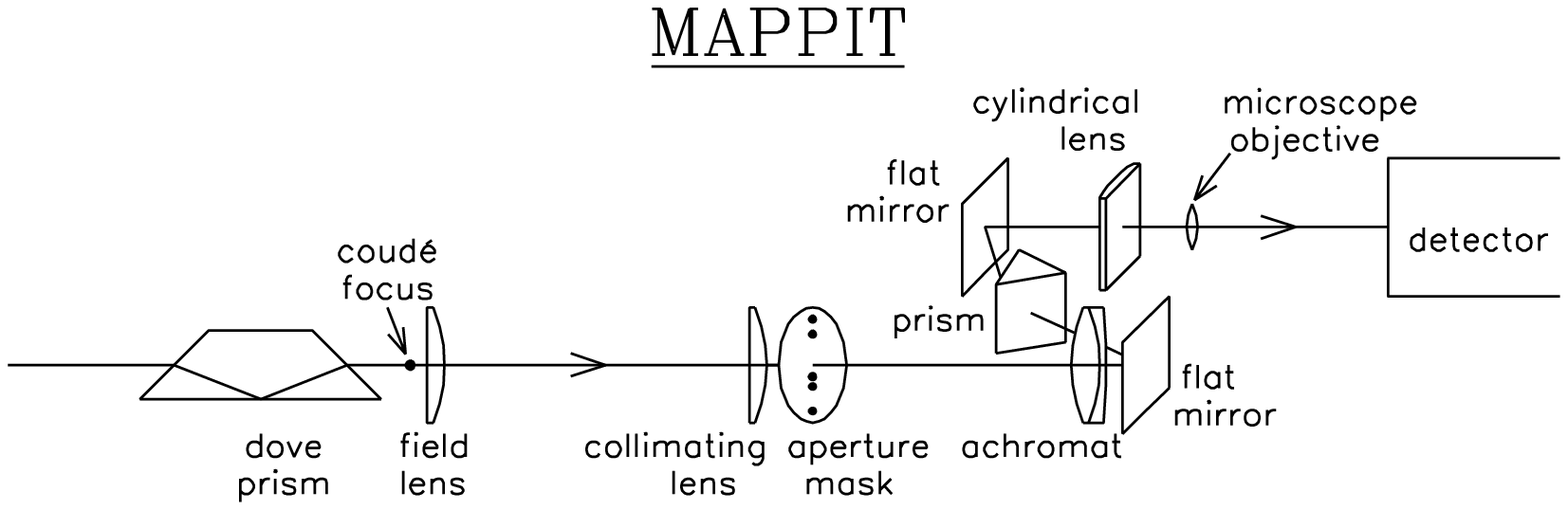}
\hfill
\includegraphics[bb=2 20 550
600,clip=true,scale=0.18]{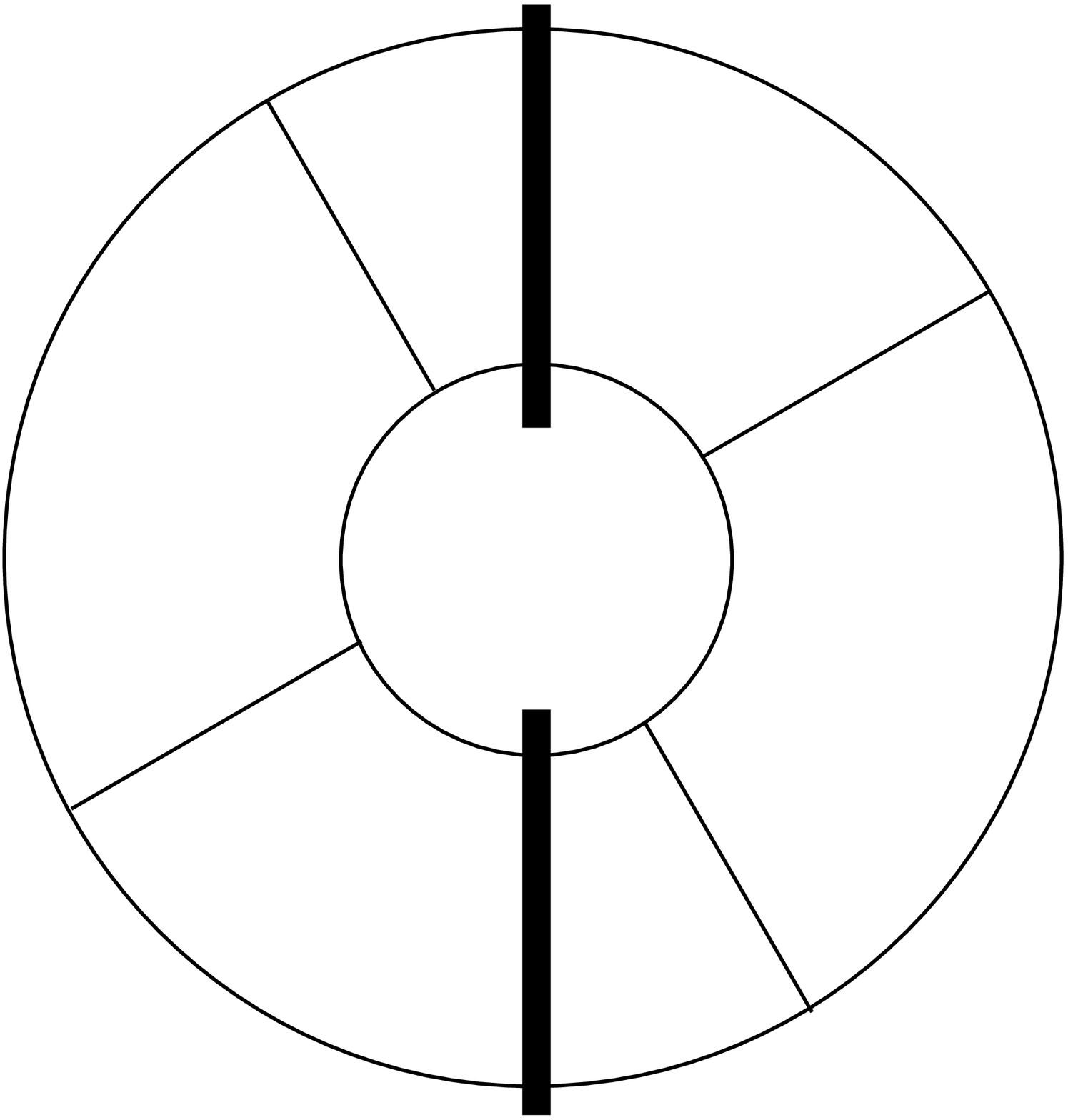} 
\caption[]{\label{mappit} A schematic view of the main components of
MAPPIT, as used in this experiment (not to scale).  Note that the field and
collimating lenses are actually achromats.  The detector was a Thomson
1024$\times$1024 CCD.  The aperture mask for the observations reported here
was a slit, whose position relative to the primary mirror of the telescope
is shown on the right.  }
\end{figure}

MAPPIT is located in the west coud\'{e} room of the AAT at Siding Spring
Observatory, Australia.  It consists of a series of optical components
mounted on two parallel optical rails (see Fig.~\ref{mappit}).  For this
experiment, the aperture mask was a slit placed diametrically across the
pupil (right panel of Fig.~\ref{mappit}).  The wavelength region was
selected by adjusting the second flat mirror, to centre the desired part of
the spectrum on the detector.  The parameters of the configuration are
given in Table~\ref{dec96params}.

\begin{table}
\caption[]{\label{dec96params} MAPPIT parameters for December 1996
observations}
\vspace{0.5cm}
\begin{center}
\begin{tabular}{lc}
\hline
Pupil diameter at aperture mask           &  25.4\,mm \\
Projected slit width    & 97\,mm          \\
Projected maximum baseline        & 3.89\,m    \\
Position angles         & 10.4$^\circ$ and 135$^\circ$      \\
Wavelength range        & 652 to 987\,nm  \\
Wavelength resolution   & 5 to 12\,nm \\
Detector pixel size                      &  19\,$\mu$m \\
Detector angular scale                   &  10.5\,mas/pixel \\
CCD window ---\\
~~spatial direction:
		& 400 pixels (not binned)             \\
~~wavelength direction:
		& 220 pixels, binned by a factor of 5 to 44 pixels   \\
Exposure time	& 10\,ms (at 650\,ms intervals)  \\
\hline
\end{tabular}
\end {center}
\label{obsparams}
\end {table}

Most previous aperture-masking observations have used a CCD with full
on-chip binning and no shutter, to collapse the interference pattern along
the fringes and produce a high-speed one-dimensional
readout\cite{BHB90,WBB92,THB97,BZvdL97}.  However, the curved fringes in a
broadband wavelength-dispersed system require full two-dimensional
detection.  For previous observations with MAPPIT in wavelength-dispersed
mode, we therefore used the IPCS (Image Photon Counting
System)\cite{Bed92,BRM94b}.  The IPCS had fast two-dimensional readout with
no readout noise but, like other photon counting systems, it had poor red
sensitivity and suffered from non-linear effects at high count rates.


The observations reported here used a CCD operating as a conventional
two-dimensional detector.  Only a portion of the CCD was read out and some
on-chip binning was used (see Table~\ref{dec96params}), to give faster
readout and reduce the effect of readout noise.  A shutter was used to set
the exposure time to 10\,ms, with one frame being recorded every 650\,ms.
The wavelength resolution ranged from about 5\,nm per binned pixel at
650\,nm to about 11\,nm per binned pixel at 950\,nm.



\section{Observations}
\label{rawdata}

R~Doradus (HR~1492; $V=5.4$; spectral type M8\,IIIe) is a nearby Mira-like
star which alternates on a timescale of several years between two pulsation
modes having periods of 332 and 175 days\cite{BZJ98}.  
Wing\cite{Win71} has pointed out that R~Dor has the brightest visual
magnitude of any M8 star and that it rivals $\alpha$~Ori as the brightest
star in the night sky at the $K$ and $L$ infrared bands.  He predicted that
it should have one of the largest angular diameters of any star.

We have previously used aperture-masking to observe R~Dor at 1.25\,$\mu$m
and found it to have an angular diameter of 57$\pm$5\,mas\cite{BZvdL97}.
This exceeds that of $\alpha$~Ori (44\,mas\cite{DBR92}) and confirms Wing's
prediction.  We detected non-zero closure phases at 855\,nm, suggesting an
asymmetric brightness distribution, but limited position-angle coverage
prevented image reconstruction.  We have presented visibility profiles for
R~Dor at 820/18\,nm (pseudo-continuum) and 850/20\,nm (TiO) which showed
that the TiO band diameter was about 20\% greater than the pseudo-continuum
diameter\cite{JBR97}.  We also found that a uniform disk gave a poor fit to
the observations.




The observations of R~Dor reported here were made on 19 December 1996
(JD~2450439).  The pulsation phase of the star at that time was
approximately 0.7 in its 332-day cycle, although the irregularity of the
pulsations makes this value uncertain\cite{BZJ98}.  Our observations
consisted of a series of 10-ms exposures of R~Dor, interspersed with
similar observations of a calibrator star ($\gamma$~Ret; HR~1264; $V=4.5$;
M4~III) and of blank sky.  The observing sequence was: calibrator (300
frames), sky (100 frames), R~Dor (300 frames), sky (100 frames).  Two such
sequences were recorded with the aperture mask at a position angle on the
sky of 10.4$^\circ$ (measured from North through East), followed by two
more sequences at position angle 135$^\circ$.  We therefore have four
separate data sets on R~Dor, two at each position angle, acquired over
about one hour.

\newpage

\begin{figure}
\centerline{\includegraphics[draft=false,width=\the\hsize]{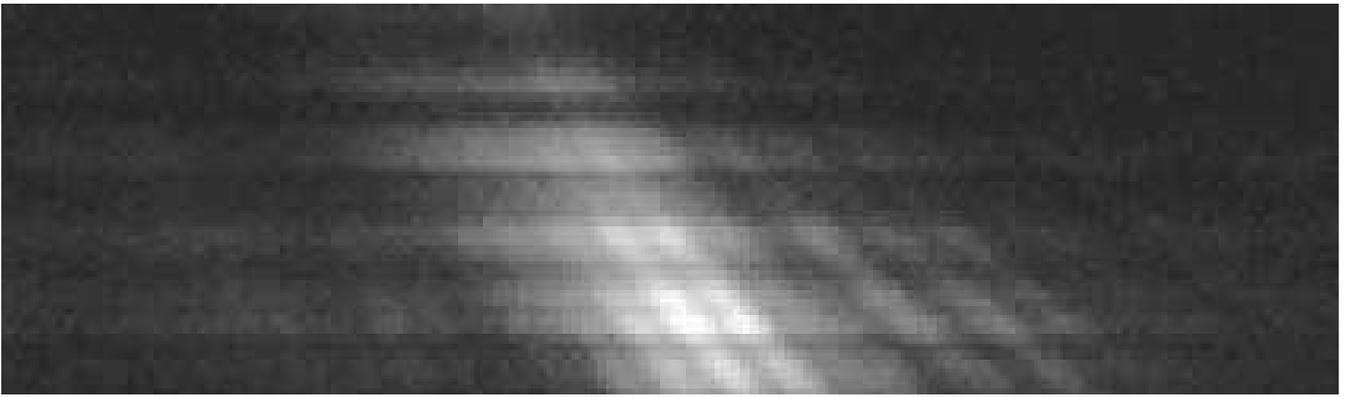}} 
\caption[]{\label{RDfringes} A single raw exposure of R~Dor.  Wavelength
increases downwards from 652\,nm to 987\,nm.  }
\end{figure}


\section{Data Processing}
\label{datared}

The data were reduced using AIPS++ (version 0.9).  The raw data consisted
of a stack of 10-ms images, each 400 $\times$ 44 pixels (spatial position
$\times$ wavelength channel).  Figure~\ref{RDfringes} shows a single frame
of data for R~Dor.  Wavelength increases downwards in 44 channels, which
are not equal in bandwidth because dispersion in the glass prism is greater
in the blue than the red.  The dark horizontal bands are TiO absorption
features in the spectrum of the star, and speckle patterns can be seen
across the image.  The prominent fringes result from inteference between
the two halves of the slit aperture (see the right panel of
Fig.~\ref{mappit}).  The fringes are diagonal rather than vertical because
of dispersion in the atmosphere and the optics, which make optical path
differences vary with wavelength.

The sky frames were averaged to a produce CCD bias frame, which was
subtracted from each source and calibrator image.  The spatial power
spectrum of each frame was calculated row by row for the 44 wavelength
channels using the Fast Fourier Transform, with the images having first
being padded with zeroes in the spatial direction to 1600 pixels and
smoothed with a Hamming window.

\begin{figure}
\bigskip
\centerline{
\hfill
\includegraphics[draft=false,width=0.47\textwidth]{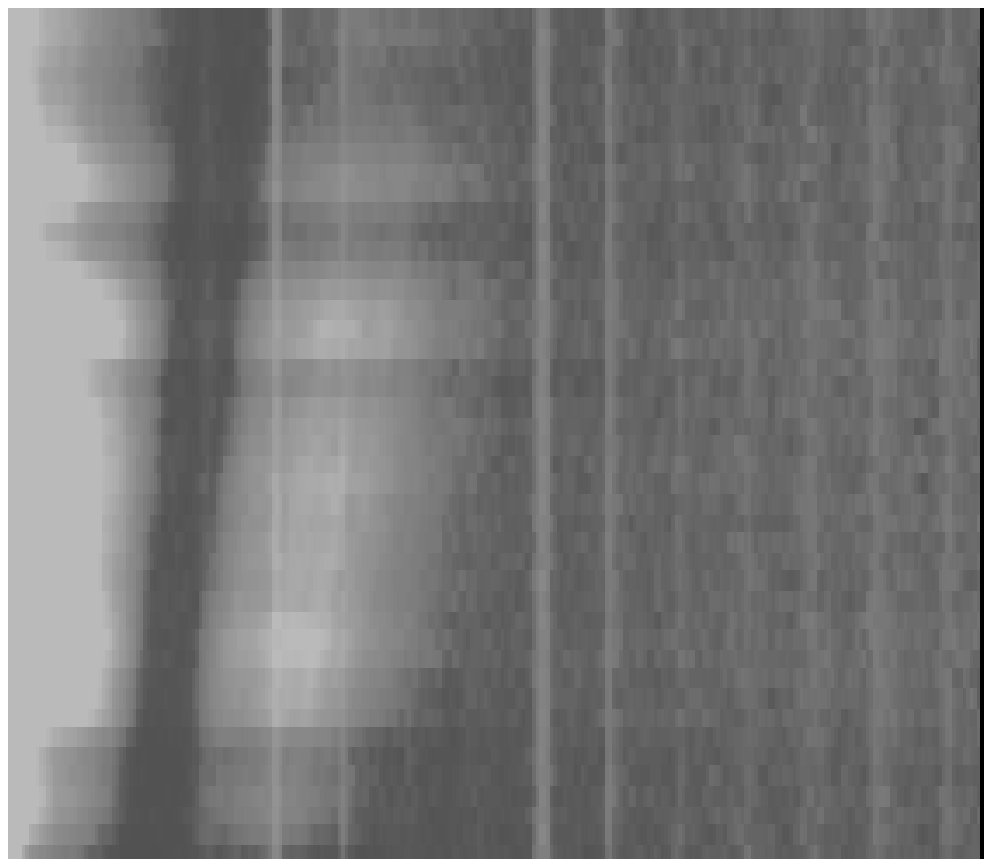}
\hfill
\includegraphics[draft=false,width=0.47\textwidth]{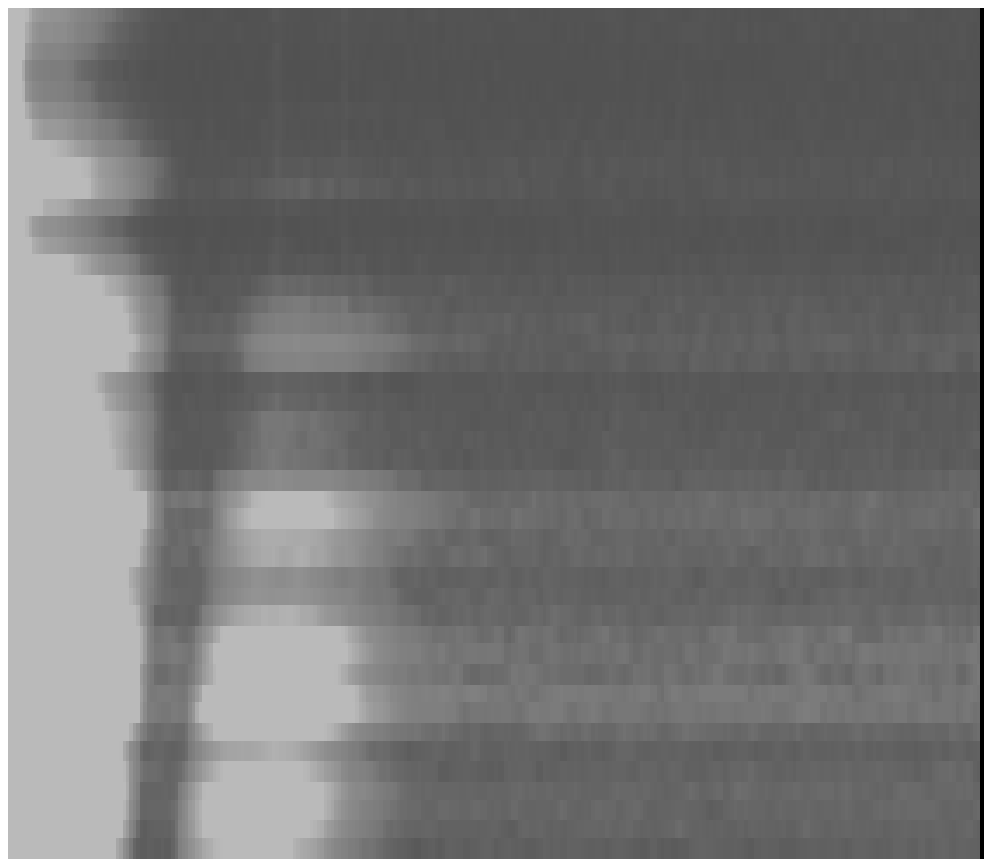}
\hfill
}
\caption[]{\label{powerspectra} Typical power spectra averaged over 100
frames for the calibrator ($\gamma$~Ret; left) and R~Dor (right).
Wavelength increases downwards, in 44 non-equal channels from 652\,nm to
987\,nm.  The horizontal axis is spatial frequency, increasing rightwards
from zero to the detector Nyquist frequency of $9.8\times10^6$\,rad$^{-1}$.
}
\end{figure}

Figure~\ref{powerspectra} shows power spectra for the calibrator star and
R~Dor, each averaged over 100 frames.  Power can be seen out to the
diffraction limit, with dark nearly-vertical bands indicating spatial
frequencies not sampled because of the telescope's central obstruction.
TiO absorption features in the stellar spectrum are visible as dark
horizontal bands.  The thin bright vertical stripes in the calibrator data
(and actually present but less obvious in R~Dor) are due to pattern noise
in the CCD electronics.  The pattern noise varied in intensity during the
observations but remained at fixed spatial frequencies, which made it
simple to ignore those regions of the power spectra.

The power spectra had a background level due to biases from photon noise
and CCD readout electronics.  This was corrected by subtracting a slowly
varying function (a cosine curve) that was fitted to regions devoid of
stellar power, namely frequencies within the central obstruction gap and
beyond the telescope diffraction limit.  The central obstruction therefore
turned out to be useful -- it is an advantage of masked-aperture
interferometry over full-aperture speckle that some regions of the power
spectrum do not contain stellar signal and are therefore available to
measure detector bias.

The next step was to turn the power estimates into visibilities, which we
did by dividing the power in each wavelength channel by the number of
photons in that channel.  We then extracted the visibility as a function of
wavelength channel at two spatial frequencies, $f_1 = 0.97\times10^6$ and
$f_2 = 3.06\times10^6$~rad$^{-1}$, chosen to avoid the central obstruction
and the CCD pattern noise.  The visibility in each wavelength channel was
averaged over a range of spatial frequencies having width
$0.37\times10^6$\,rad$^{-1}$.  

Note that the baseline corresponding to a given spatial frequency is
proportional to wavelength.  Over the range 650--950\,nm, spatial
frequency~$f_1$ corresponds to baselines 0.63--0.92\,m, while $f_2$
corresponds to 1.99--2.91\,m.


\begin{figure}
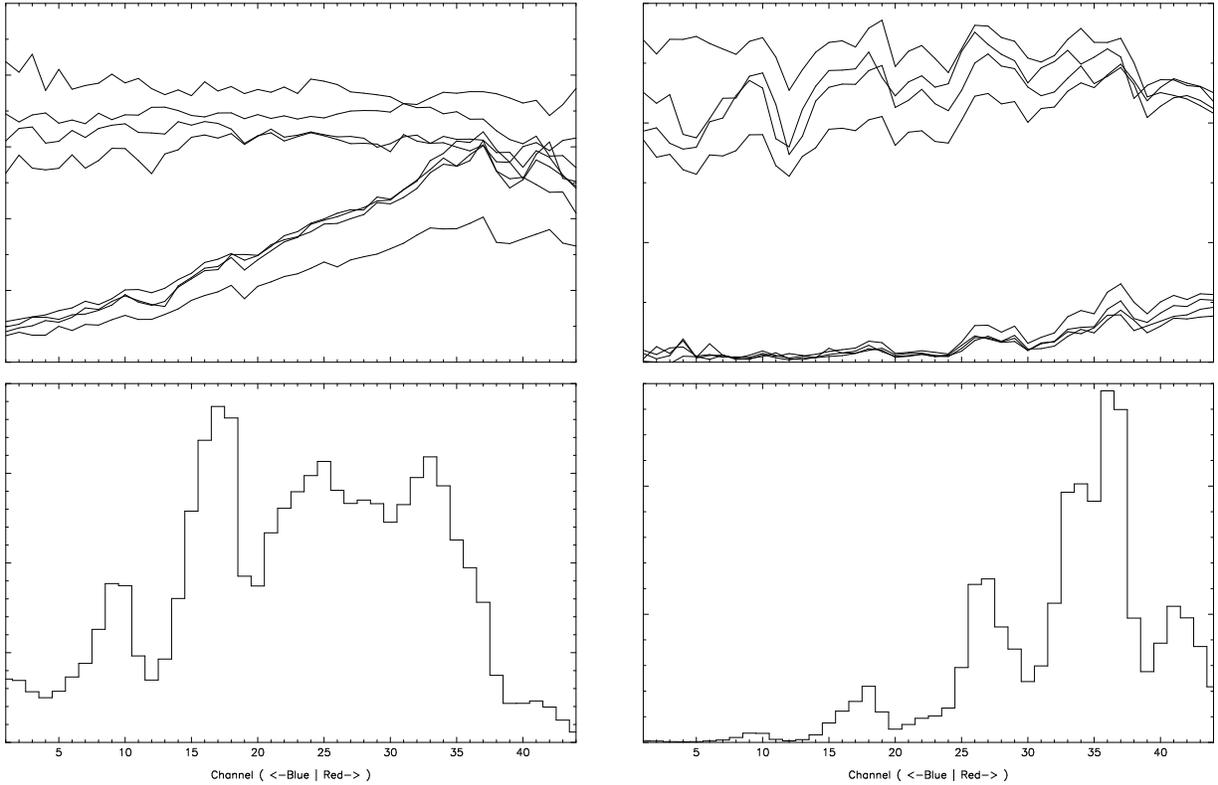

\centerline{
\hfill
\includegraphics[angle=-90,bb=60 70 520 770, clip=true,
width=0.43\textwidth]{p86b.ps}
\hfill
\includegraphics[angle=-90,bb=60 70 520 770, clip=true,
width=0.43\textwidth]{p86c.ps}
\hfill
}
\centerline{
\hfill
\includegraphics[angle=-90,bb=60 70 570 770, clip=true,
width=0.43\textwidth]{p86a.ps}
\hfill
\includegraphics[angle=-90,bb=60 70 570 770, clip=true,
width=0.43\textwidth]{p66bb1.ps}
\hfill
}
\caption[]{\label{p86a} Uncalibrated power (visibility squared) versus
wavelength channel for the calibrator (upper left) and R~Dor (upper right).
Results are shown at two spatial frequencies for four observations (two at
each position angle).  The lower panels show the observed spectrum of each
star (uncorrected for the CCD response).}
\end{figure}

The results for both calibrator and R~Dor are shown in Fig.~\ref{p86a}.
The squared visibilities (upper panels) have not yet been corrected for the
atmospheric transfer function.  For each star, the upper four curves
correspond to $f_1$ while the lower curves correspond to $f_2$ (two
observations at two position angles -- see Sec.~\ref{rawdata}).  The lower
panels show the stellar spectra (not yet corrected for the CCD response
function), which indicate the positions of the TiO absorption bands.

A number of features can be seen in Fig.~\ref{p86a}.  The offsets between
the four curves at each spatial frequency are due to variations in the
seeing.  In both stars, the visibilities at $f_2$ drop steeply towards the
blue, presumably due to wavelength dependence of the atmospheric transfer
function.  The visibilities at $f_1$, on the other hand, show little
evidence of such a trend.

The purpose of these observations was to measure the variation of
visibility with wavelength due to the structure of the star's atmosphere.
In R~Dor we certainly see strong variations which coincide with absorption
features in the stellar spectrum.  In the case of the calibrator
($\gamma$~Ret; spectral type M4\,III), the TiO bands are present in the
spectrum but weaker and we do not expect them to be accompanied by
significant changes in visibility\cite{H+S98}.  This seems to be the case,
although we do see small drops in visibiity near wavelength channels 12, 19
and 38.  These probably indicate residual systematic effects arising from
inaccurate background subtraction.




The strong visibility decreases shown by R~Dor in the TiO bands, and the
lack of similar variations for $\gamma$~Ret, gives us confidence that these
variations are intrinsic to R~Dor.  There is also a clear consistency in
the change in visibility from observation to observation and between the
two spatial frequencies.

\begin{figure}
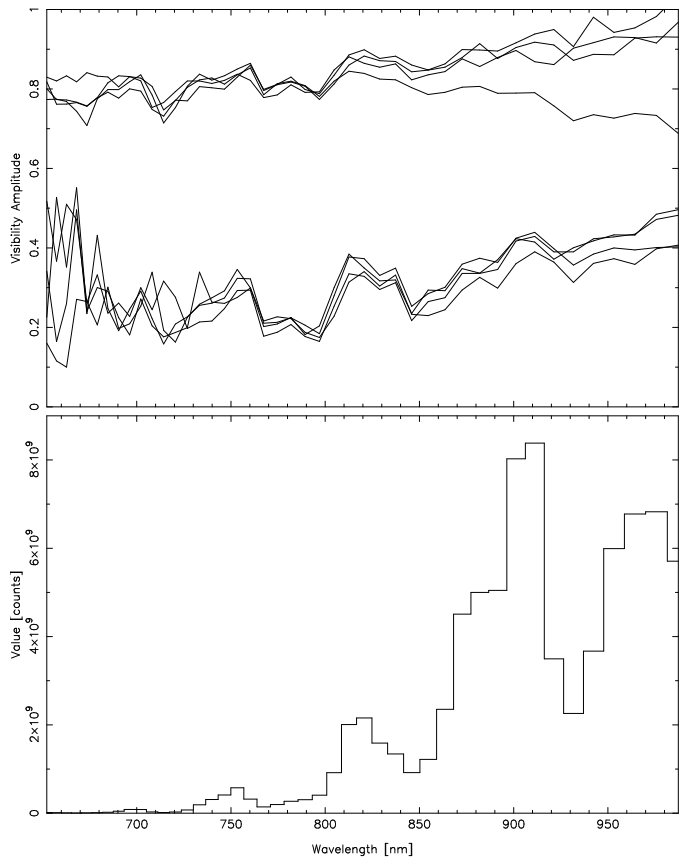

\centerline{\includegraphics[angle=-90,bb=76 34 520 769, clip=true,
width=0.5\textwidth]{p78ee.ps}}
\centerline{\includegraphics[angle=-90,bb=76 34 570 769, clip=true,
width=0.5\textwidth]{p78cc.ps}}
\caption[]{\label{p78eecc} The calibrated visibility (top) and flux
(bottom) for R~Dor, as a function of wavelength.  As in Fig.~\ref{p86a},
visibilities are shown at two spatial frequencies for four observations
(two at each position angle), but normalized as described in the text.  }
\end{figure}

The visibilities for R~Dor were calibrated using those of the calibrator
star, to remove any instrumental and atmospheric effects that were common
to both.  Note that we expect the calibrator star $\gamma$~Ret to be
slightly resolved.  This star has no measured angular diameter, but
Ochsenbein et al.\ estimated a diameter of 11\,mas based on its spectral
type\cite{O+H82}, while the $V-K$ calibration of Di~Benedetto gives
7.5\,mas\cite{DiB93}.  We have adopted an intermediate value of 9.25\,mas,
which we used to correct the measured visibility spectra.  We then
calibrated each of the four R~Dor observations by dividing the visibilities
by those of the corresponding calibrator observations.  The results are
shown in Fig.~\ref{p78eecc}, where the horizontal axis has now been
converted to wavelength and the spectrum of the star has been approximately
corrected for the quantum efficiency curve of the CCD\@.

Note that even after calibration, there remained systematic offsets up to
about 10\% between the four visibility curves.  There was no obvious
dependence on position angle, such as would arise from a non-circular
stellar disk, so we attribute these offsets to residual effects of seeing.
Since we are primarily interested in the variation of visibility with
wavelength, these offsets have been removed in Fig.~\ref{p78eecc} by
normalizing so that the average value at $f_1$ in wavelength channels~1
through~34 is the same for all four observations.  Note that the
measurements at spatial frequency $f_2$ become noisy at the blue end due to
the low visibility and low photon flux.





\section{Comparison with stellar models}

It has been pointed out in the literature that improvements in atmospheric
models for M giants requires measurements of stellar radii over a wide
range of wavelengths \cite{Sch85,BBS89a}.  The visibility curves in
Fig.~\ref{p78eecc} represent the first multi-wavelength visibility
measurements of a red giant star.  We have compared our observations of
R~Dor with the M-giant models by Bessell, Brett, Hofmann, Scholz and
Wood\cite{BBS89b,BBS91,BSW96,HSW98}.  Note that although most of their
dynamic models were originally developed for $o$~Cet, they are
representative of typical M-type Miras and, in some cases, have been used
in previous studies for these types of stars.

We find agreement in the general features, with clear drops in visibility
within the TiO absorption bands.  However, although some models match the
visibility observations in some regions, none match across the whole
spectrum.  Also, most of the models overpredict the ratio between
visibility amplitudes at spatial frequencies~$f_1$ and~$f_2$.  The static
1M$_{\odot}$ models, in particular, did not fit our observations well.

\begin{figure}
\centerline{\includegraphics[bb=74 174 505 661,
width=0.58\textwidth]{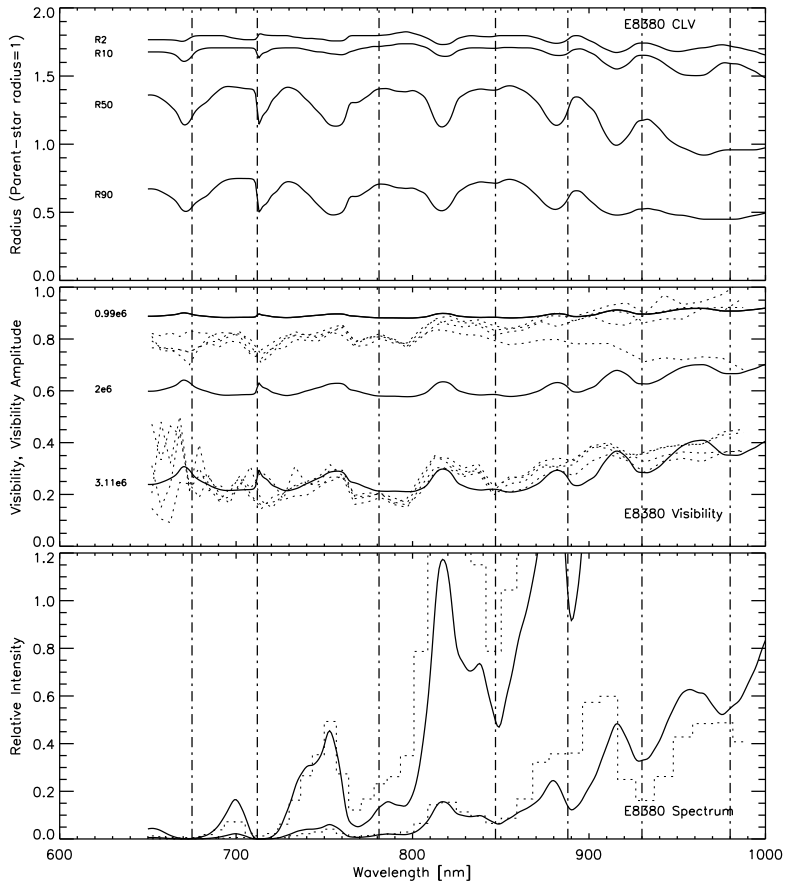}}
\caption[]{\label{p84d} Comparison of observed visibilities of R~Dor
(dotted curves, reproduced from Fig.~\ref{p78eecc}) with a 1M$_{\odot}$
overtone-mode model (E8380).  The Rosseland angular diameter of the model
was set to 49\,mas.  The top panel shows the model radius at 2, 10, 50 and
90\,\% of the central intensity.  The middle panel shows model visibilities
at three spatial frequencies ($f_1$, $f_2$ and an intermediate one).  The
bottom panel shows the flux spectrum for the model and observations,
both plotted twice with vertical scales differing by a factor of 20 to
accommodate the large range in intensities.  To guide the eye, vertical
dashed lines indicate the strongest absorption bands. }
\end{figure}

\begin{figure}
\centerline{\includegraphics[bb=74 174 505 661,
width=0.58\textwidth]{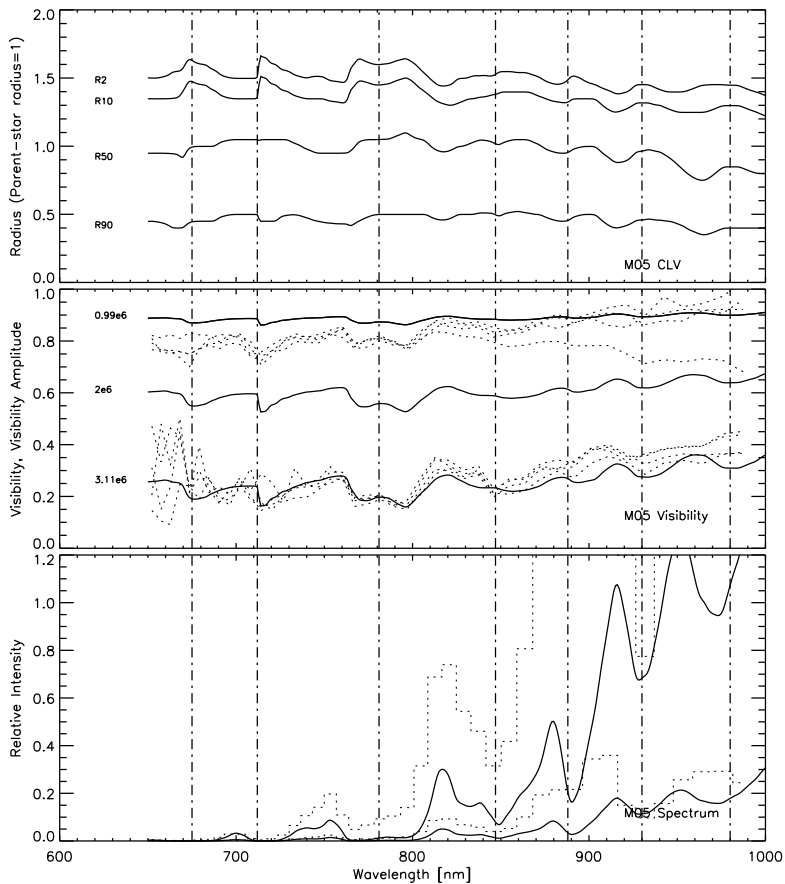}}
\caption[]{\label{p84ihere} Same as Fig.~\ref{p84d} for the 1.2M$_{\odot}$
fundamental-mode model (M96400), with a Rosseland diameter of 53\,mas.}
\end{figure}

Figs.~\ref{p84d} and \ref{p84ihere} show our observations superimposed on
visibility spectra calculated for two Mira models: a 1M$_{\odot}$ star
pulsating in the first overtone and a 1.2M$_{\odot}$ star in the
fundamental mode.  The model calculations are described in full by Jacob et
al\cite{JBR2000}.  The model data were smoothed to simulate observing with
a filter of 10\,nm FWHM, while the spectral resolution of the observations
varies from 5 to 12\,nm (see Sec.~\ref{sec.mappit}).  The radius of each model
was adjusted until the visibilities at $f_2$ were approximately aligned.
The corresponding angular diameters for the models are given in the figure
captions.

The overtone-mode 1M$_{\odot}$ model (Fig.~\ref{p84d}) does not fit the
observations very well, depsite the fact that it was chosen by Jacob et
al.\cite{JBR2000} because it had parameters close to those of R~Dor.  The
predicted variation in visibility with wavelength is too small, especially
at spatial frequency~$f_1$.  The model also predicts an {\em increase\/} in
visibility within the 710-nm TiO band that is not observed (this extreme
limb darkening effect is discussed by Jacob et al.\cite{JBR2000}).  Other
overtone-mode 1M$_{\odot}$ models at other phases also give poor fits to
the observations.

The fundamental-mode 1.2M$_{\odot}$ model at a pulsation phase of 0.5
(Fig.~\ref{p84ihere}) gives a much better fit, reproducing many of the
features in the lower ($f_2$) visibility curve.  Visibility variations at
$f_1$ are present in the model, although still less pronounced than in the
observations.  Note that the same model at maximum phase is a poor fit to
the observations.


In summary, the 1.2M$_{\odot}$ M-series models (and 1M$_{\odot}$ D models)
appear to be the best fits to our observations.  Both these models pulsate
in the fundamental mode.  The first-overtone-mode models (E series) do not
fit well, perhaps because they are at the wrong pulsation phase (1.0, 1.1
\& 1.21).

\section{Conclusion}

We have presented multi-wavelength visibility observations of R~Dor and
compared them with theoretical models.  We find that fundamental-mode
1M$_{\odot}$ and 1.2M$_{\odot}$ Mira models provide the best fits to our
observations.  First overtone 1M$_{\odot}$ Mira models and static models do
not fit our observations well.  Therefore, if R~Dor does pulsate in the
first overtone mode (as Bedding et al.\cite{BZJ98} suggest) then the
first-overtone models will need to be extended to other parameters and
phases, and/or to be refined to accommodate our results.

Finally, we note that there is still uncertainty as to whether the
pulsation mode of Miras is the fundamental or the first overtone.  Observed
line velocities are too large for first-overtone pulsation, but angular
diameters are too large for fundamental-mode pulsation.  To resolve this,
it has been suggested that scattering, e.g., by molecules or dust, may
cause angular diameter measurements to over-estimate the true stellar
size\cite{PCDFR99}.  Multi-wavelength observations like those presented
here should be able to place constraints on the amount of scattering that
can be present.

\acknowledgments

We thank the staff of the Anglo-Australian Observatory for their invaluable
support.  The development of MAPPIT was supported by a grant under the
CSIRO Collaborative Program in Information Technology, and by funds from
the University of Sydney Research Grants Scheme and the Australian Research
Council.  CAH is grateful to the Royal Society for financial support.

  \end{document}